\documentclass[twocolumn,preprintnumbers,gensymb,amsmath,amssymb,aps,prb,superscriptaddress]{revtex4}
\usepackage{graphicx}

\begin{document}

\title{Structural Transitions in Vortex Systems with Anisotropic Interactions } 

\author{M.W. Olszewski}
\affiliation{Theoretical Division, Los Alamos National Laboratory, Los Alamos, New Mexico 87545 USA}
\affiliation{Department of Physics, University of Notre Dame, Notre Dame, Indiana 46656, USA}

\author{M.~R.~Eskildsen}
\affiliation{Department of Physics, University of Notre Dame, Notre Dame, Indiana 46656, USA}

\author{C. Reichhardt}
\affiliation{Theoretical Division, Los Alamos National Laboratory, Los Alamos, New Mexico 87545 USA}

\author{C.J.O. Reichhardt}
\affiliation{Theoretical Division, Los Alamos National Laboratory, Los Alamos, New Mexico 87545 USA}

\date{\today}

\begin{abstract}
  We introduce a model of vortices in type-II superconductors with a
  four-fold anisotropy in the vortex-vortex interaction potential.
Using numerical simulations we show that the vortex lattice undergoes structural transitions as the anisotropy is increased,
with a triangular lattice at low anisotropy, a rhombic intermediate state, and a square lattice for high anisotropy.
In some cases we observe a multi-$q$
state consisting of an Archimedean tiling
that combines square and triangular local ordering.
At very high anisotropy,  domains of vortex chain states appear. 
We discuss how this model can be generalized to higher order anisotropy as well as its applicability to other particle-based systems with anisotropic particle-particle interactions.   
\end{abstract}

\maketitle

\section{Introduction}
A type-II superconductor subjected to a magnetic field
forms vortices that each carry one quantum of magnetic flux.
Due to their mutual repulsion,
these vortices arrange themselves in an ordered vortex lattice (VL)
the vortex-vortex interactions dominate external influences
such as pinning by impurities or thermal disordering.
The VL is triangular in isotropic superconductors
\cite{1};  however, anisotropy in the vortex-vortex interactions can cause
changes in the VL symmetry.\cite{2}
One of the most studied systems exhibiting a transition to a square
VL is borocarbide materials \cite{3,4,5,6,7,8,9,10}. 
Transitions to square lattices have also been observed in high temperature superconductors,\cite{11,12,13,14}
heavy fermion materials,\cite{15,16,17,18}
and spin triplet superconductors such as Sr$_2$RuO$_4$.\cite{19,MRE1,MRE2}
Additionally, transitions from triangular to square VLs
can arise in other superfluid systems including Bose-Einstein condensates \cite{20}
and dense nuclear matter in extreme conditions.\cite{21}

\begin{figure*}
\includegraphics[width=5.0in]{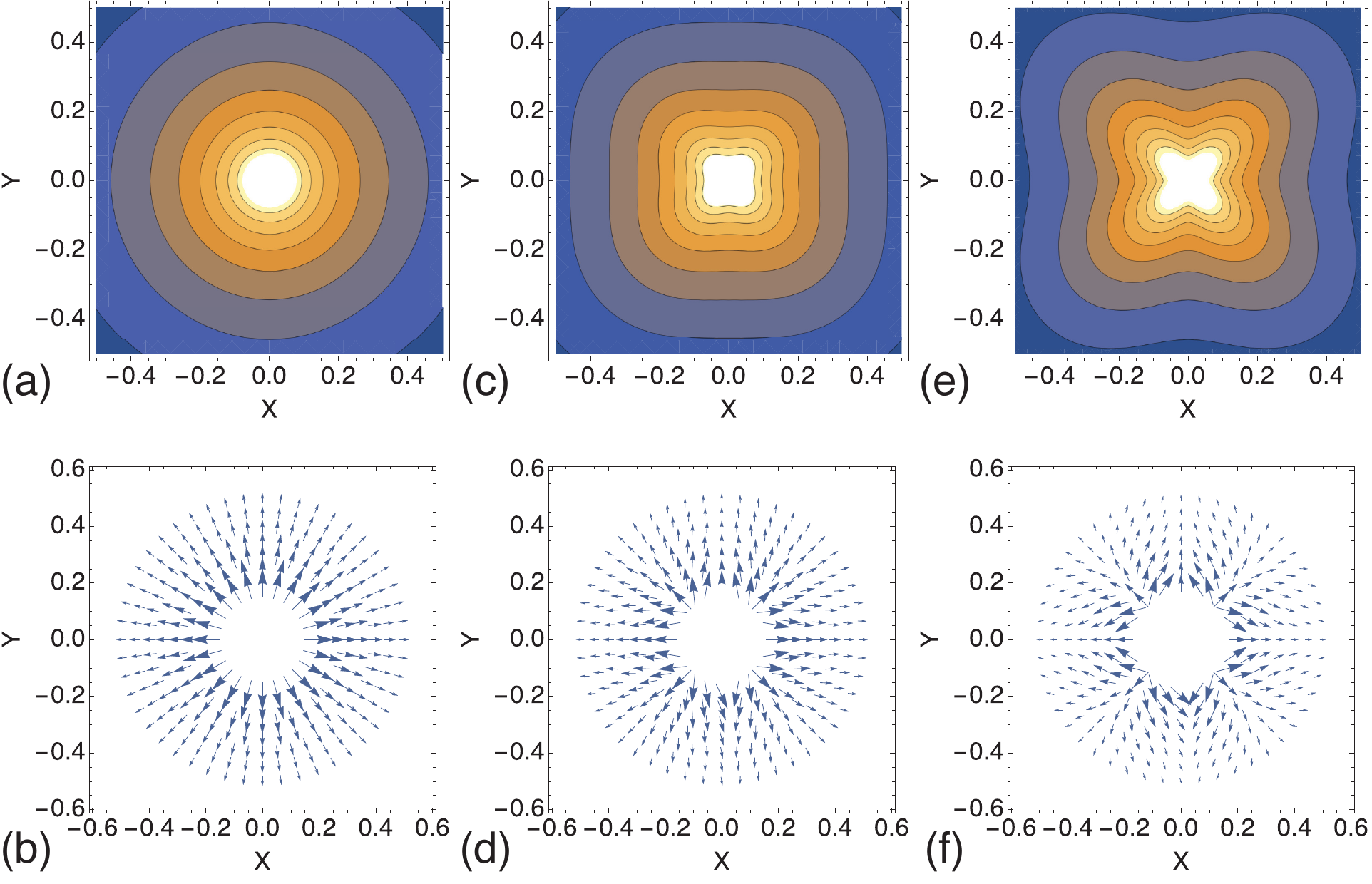}
\caption{
  Equipotential lines (a,c,e) and force fields (b,d,f)
  for the vortex-vortex interaction potential in Eq. 1 with $n_a=4$
  and $\phi = 45^{\circ}$.
  (a,b) Isotropic case with anisotropy strength
  $A_a = 0$.
  (c,d) At $A_a = 0.1$, nonradial forces begin to appear.
  (e,f) At $A_a = 0.25$ the nonradial forces are stronger.
}
\label{fig:1}
\end{figure*}

In superconducting systems, theoretical approaches used to study the
VL structural transitions include
modifications to the London model \cite{22,MRE3},
addition of four-fold symmetric terms to the Ginzburg-Landau free energy \cite{23},
Eilenberger theory \cite{24},
modified Ginzburg-Landau approaches \cite{25},
and modifications to the vortex interactions produced by strain fields \cite{26}.
Notably absent
from this list
are molecular dynamics (MD) simulations,
which treat the vortices as point particles with bulk Bessel function interactions
or thin film Pearl interactions.
Until now, MD methods have only been applied to isotropic
pairwise potentials, which produce a triangular VL
in clean systems \cite{27,28,29,30,31,N,32,33,34,35,36}.
One of the issues is that simply adding a stronger radial repulsive force along certain  directions does not capture the triangular to square transitions
in the vortex system.
Instead, it is essential to consider the full anisotropic potential in which nonradial forces also arise.
In addition to the equilibrium VL configurations, MD simulations
give access to the statics and dynamics of a large number of vortices over long times. 

Here we introduce a model for vortices in anisotropic superconductors where
the vortex-vortex interaction potential is extended to include a fourfold anisotropy.
Using MD simulations we show that as the anisotropy is increased,
the VL symmetry changes from triangular to rhombic to square.
Additionally, in some cases we find a multi$-q$
lattice state consisting of an Archimedean tiling with a combination of square and triangular local ordering.
For very large anisotropy, domains of vortex chains appear.
Our model can be generalized to any anisotropy and provides a basis for modeling
VL reorientation
or other vortex symmetry transitions  such as those observed in
multigap superconductors.\cite{N2}
The model can be applied not only to vortices in type-II superconductors, but also to
other particle-based systems with anisotropic interactions where triangular to square transitions can arise.
These include skyrmion lattices, 
where triangular to square transitions have recently been observed,\cite{37,38}
as well as colloidal particles with anisotropic interactions.\cite{39,40}

\section{Model}
In an isotropic bulk superconductor, the vortex-vortex interaction potential is typically isotropically repulsive and takes the form of
a zeroth order Bessel function, $U(r) = K_{0}(r)$.\cite{N3}
For anisotropic materials, we propose the following modified pair potential for the vortex interactions:
\begin{equation}
  U(r) = A_{v} K_{0}(r) \left[ 1 + A_a \cos^2 \left( \frac{n_a (\theta - \phi)}{2} \right) \right]
\end{equation}
where
$r=|{\bf r}_i-{\bf r}_j|$ is the distance between two vortices at positions ${\bf r}_i$ and ${\bf r}_j$.
The angle between the two vortices with respect to the positive $x$ axis is given by
$\theta=\tan^{-1}(r_y/r_x)$
where ${\bf r}={\bf r}_i-{\bf r}_j$,
$r_x={\bf r}\cdot {\bf \hat x}$, and $r_y={\bf r}\cdot {\bf \hat y}$.
The rotation angle of the anisotropic axes is defined to be $\phi$, and $n_a$ is the order of the anisotropic interaction.
The prefactor $A_{v}$ represents the strength of the isotropic component of the vortex-vortex pair interaction force, and $A_a$
controls the amplitude of the anisotropic interaction.

To fully understand the manner in which Eq. (1) produces nonradial interactions,
we examine the nature of the anisotropic forces generated by $U(r)$.
For nonzero values of $A_a$,
the potential is stronger along certain interaction directions and weaker between these directions.
The anisotropic order $n_a$ determines the number of strong interaction directions, which are evenly spaced in $\theta$.
For example, for $n_a=6$ and $\phi = 0$,
the interaction potential passes through local maxima at $\theta=0^\circ$, $60^\circ$, $120^\circ$, $180^\circ$, $240^\circ$, and $300^\circ$.
Setting $\phi = 15^{\circ}$,
the local maxima of the interaction potential are shifted to $\theta=15^\circ$, $75^\circ$, $135^\circ$, $195^\circ$, $255^\circ$, and $315^\circ$. 
Between these values of $\theta$, the interaction strength varies as a cosine.
The force field from the potential is ${\bf F}_{vv} = -\nabla U = (-\frac{\partial U}{\partial x}, -\frac{\partial U}{\partial y})$
with components
\begin{widetext}
\begin{eqnarray}
  F_{x}
     & = & A_v \left[ \cos(\theta) K_{1}(r) \left( 1+ A_a \cos^2 \left( \frac{n_a(\theta - \phi)}{2} \right) \right) 
       - \frac{\sin(\theta)}{r} K_{0}(r) \frac{n_a A_a}{2} \sin( n_a (\theta - \phi)) \right]\\
  F_{y}
    & = & A_{v} \left[ \sin(\theta) K_{1}(r) \left( 1 + A_a \cos^2 \left( \frac{n_a(\theta - \phi)}{2} \right) \right)
        + \frac{\cos(\theta)}{r} K_{0}(r) \frac{n_a A_a}{2} \sin(n_a(\theta - \phi)) \right].
\end{eqnarray}
\end{widetext}
In the present work
we focus on the $n_a = 4$ interaction potential with fourfold anisotropy; however, by changing $n_a$ our model can be applied to twofold ($n_a = 2$), sixfold ($n_a = 6$), or other degrees of anisotropy.
In Fig.~\ref{fig:1}(a,b) we plot the
equipotential lines and force field
for $U(r)$ from Eq.~1 for the isotropic case with $A_a = 0$, where
the equipotential lines are circularly symmetric
and the forces are strictly radial.
At $A_a = 0.1$ in Fig.~\ref{fig:1}(c,d),
the fourfold symmetry is apparent and weak nonradial forces appear.
In Fig.~\ref{fig:1}(e,f) at $A_a = 0.25$, the fourfold symmetry is much more pronounced
and the nonradial forces are clearly visible.

\begin{figure}
\includegraphics[width=3.5in]{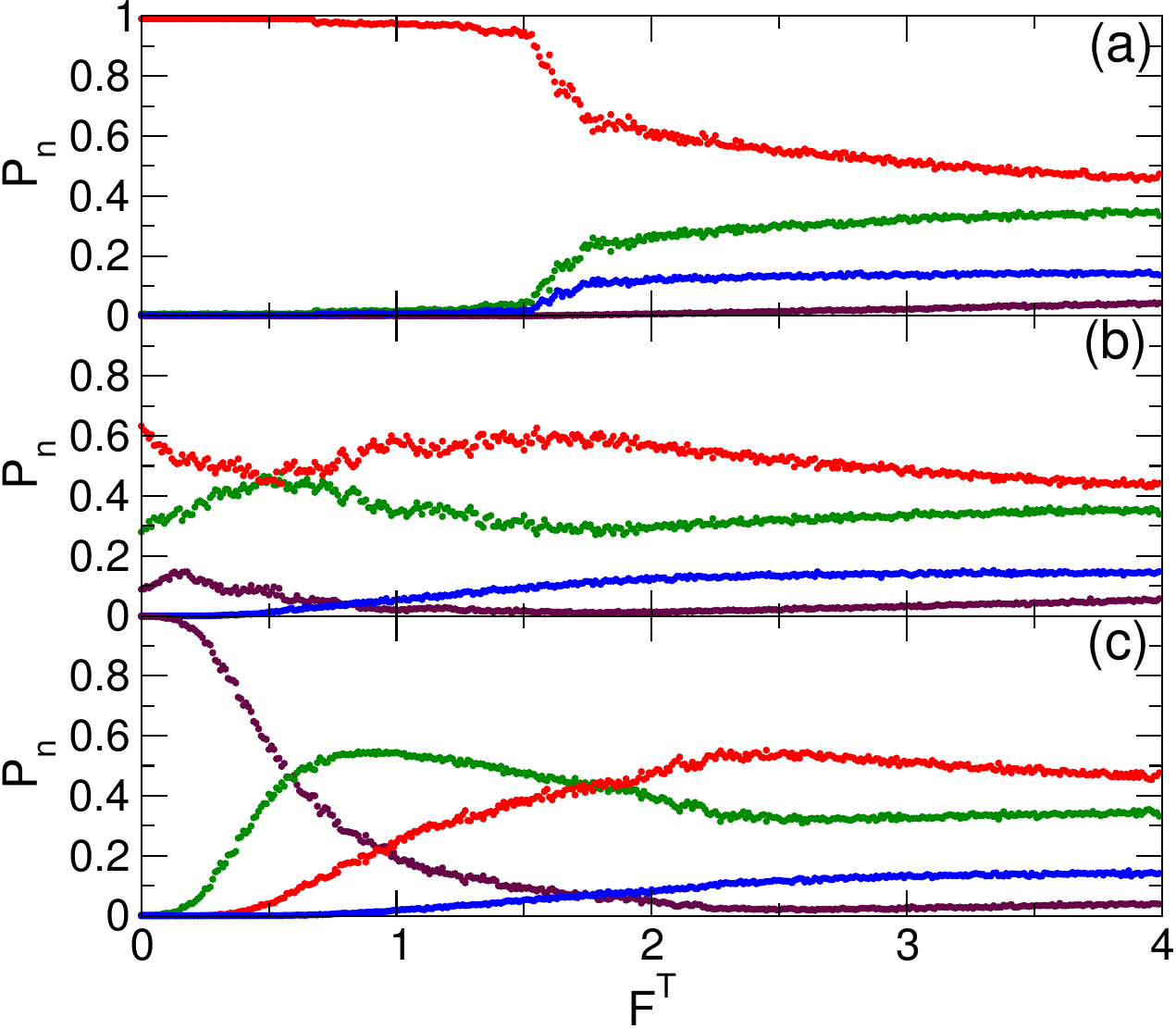}
\caption{Measures $P_n$ of VL 
              ordering versus temperature $F^T$ during the annealing procedure for
  $P_{7}$ (blue), $P_6$ (red), $P_5$ (green), and $P_4$ (maroon) in samples with $A_v = 2.0$.
  (a) In the isotropic system with $A_a = 0$, the vortices form a triangular lattice with $P_6 = 1.0$.
  (b) For $A_a = 0.039$, rhombic ordering appears.
  (d) At $A_a = 0.099$, the vortices form a square lattice with $P_4 = 1.0$.       
  }
\label{fig:2}
\end{figure}

\begin{figure}
\includegraphics[width=3.5in]{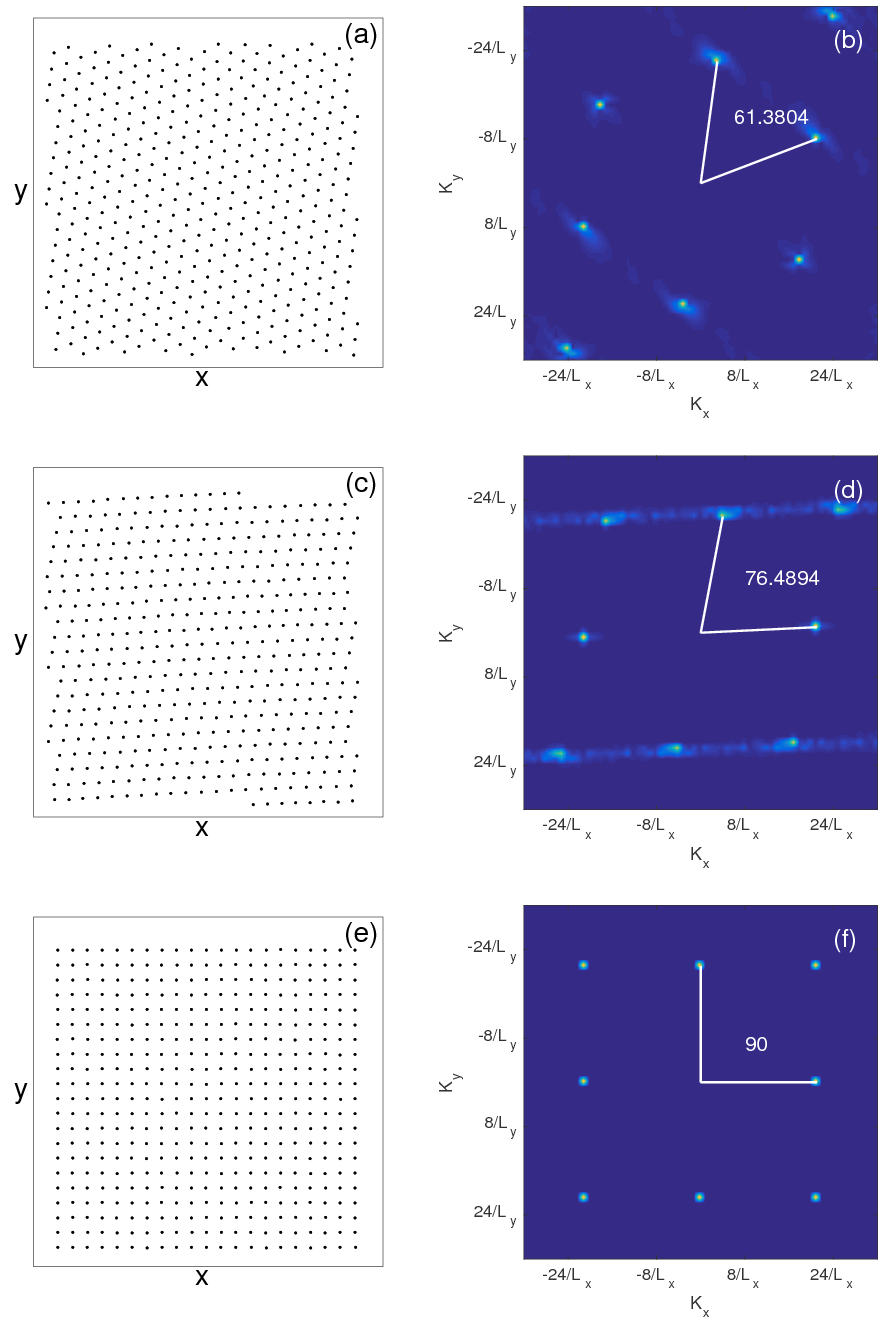}
\caption{Real space
  VL configurations (a,c,e) and corresponding heightfield plots of the structure factor $|S({\bf k})|$ (b,d,f)
  after annealing in samples with $A_v = 2.0$.
  (a,b) The isotropic system with $A_a = 0$ has triangular ordering and a lattice angle $\theta_l$ (marked in white) of $\theta_l \approx 60^{\circ}$.
  (c,d) At $A_a = 0.039$, the vortices form a rhombic lattice with $\theta_l \approx 76^{\circ}$. 
  (e,f) At $A_a = 0.099$, the vortices form a square lattice with $\theta_l=90^{\circ}$.
 }
\label{fig:3}
\end{figure}

To investigate the VL
ground states that emerge as the anisotropy of the potential increases,
we perform MD simulations of
$N=441$ vortices in a two-dimensional system of size $L \times L$ with $L=36\lambda$
and periodic boundary conditions in the $x$ and $y$ directions.
Distances are measured in units of the London penetration depth $\lambda$.
The dynamics of vortex $i$ is governed by
an overdamped equation of motion:  
\begin{equation}
\eta \frac{d {\bf r}_{i}}{dt} = {\bf F}^{i}_{vv} + {\bf F}^{i}_{T}.
\end{equation}
Here $\eta$ is the damping constant which we set equal to unity.
Thermal forces are modeled by Langevin kicks ${\bf F}_T^i$ which have the properties
$\langle {\bf F}_{T}\rangle = 0.0$ and 
$\langle {\bf F}^{i}_{T}(t){\bf F}^{j}_{T}(t^{\prime})\rangle = 2\eta k_{B}T
\delta_{ij}\delta(t - t^{\prime})$ 
where $k_{B}$ is the Boltzmann constant.
We perform simulated annealing by
starting in a high temperature molten state and gradually cooling the system to $T = 0$.
We use an initial temperature of $F^{T} = 4.0$ and
decrement
the temperature by
$\Delta F^T = -0.01$ every 40,000 simulation time steps,
which is long enough to ensure that the system reaches an equilibrium state.

\begin{figure}
\includegraphics[width=3.5in]{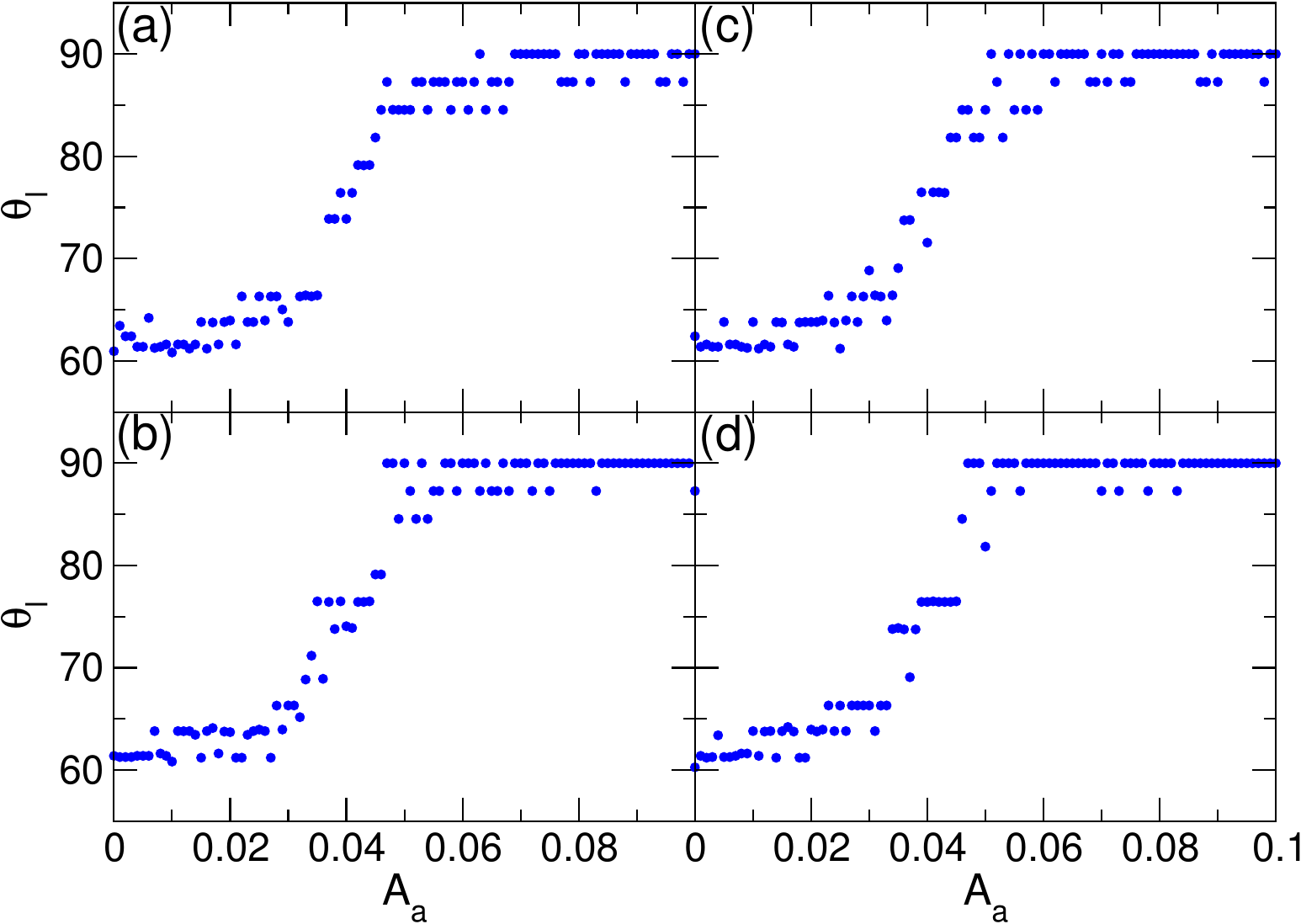}
  \caption{
    Lattice angle $\theta_l$ versus $A_a$ for systems with (a) $A_{v} = 1.0$, (b) $A_{v} = 1.5$, (c) $A_{v} = 2.0$, and (d) $A_{v} = 2.5$.
    In all cases, the vortices form a triangular VL
    at small $A_a$ with $\theta_l \approx 60^{\circ}$,
    pass through an intermediate rhombic state with $\theta_l \approx 75^{\circ}$,
    and then form a square lattice with $\theta_l \approx 90^{\circ}$ at large $A_a$.
}
\label{fig:4}
\end{figure}

\section{Vortex Structures}
We use a Voronoi construction to obtain the local coordination number $z_i$ of each vortex,
and compute the fraction $P_n$ of vortices with coordination number $n$ using $P_n=N^{-1}\sum_{i=1}^{N}\delta(z_i-n)$ for $n=4$, 5, 6, and 7.
Figure~\ref{fig:2}(a) shows
$P_4$, $P_5$, $P_6$, and $P_7$ versus $F^T$ obtained during the annealing process
in an isotropic system with $A_a = 0$ and $A_v = 2.0$.
Initially the system is in a high temperature molten state,
and as $F^T$ is reduced the vortices order into a triangular lattice with $P_6 = 1.0$.
In Fig.~\ref{fig:2}(b) at $A_a = 0.039$, the vortices freeze into a state with rhombic ordering,
while in Fig.~\ref{fig:2}(c) at $A_a = 0.099$, the vortices form a square lattice with $P_4 = 1.0$.
Together, these results show how the equilibrium
($T = 0$) value of $P_6$ is suppressed and
the value of $P_4$ grows as the anisotropy is increased.
Compared to the drop in $P_6$, the rise of $P_4$ is more gradual and happens at a lower $F^T$.

\begin{figure}
\includegraphics[width=3.5in]{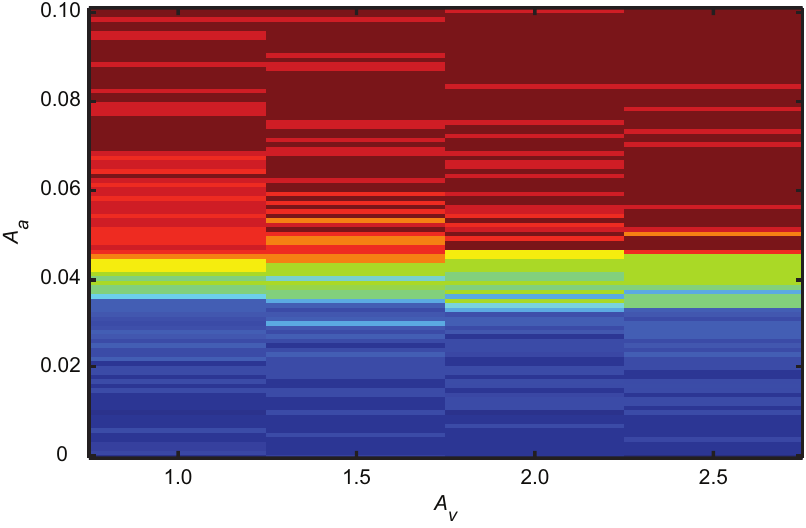}
\caption{
  Lattice ordering phase diagram as a function of $A_a$ versus $A_v$.
  Blue: triangular order; red: square order; yellow and green shades: rhombic
  order, which is centered around $A_a = 0.04$.
}
\label{fig:5}
\end{figure}

We can further characterize the final $F^T=0$ state
using the structure factor
$S({\bf k})=N^{-1}|\sum_i^{N}\exp(-i{\bf k} \cdot {\bf r}_i)|^2 $.
In Fig.~\ref{fig:3}(a)
we illustrate the final real space vortex positions in a sample with
$A_a = 0$ and $A_v = 2.0$, and
in Fig.~\ref{fig:3}(b) we plot the corresponding $|S({\bf k})|$ as a heightfield.
The lattice angle $\theta_l$ is defined to
be the angle between adjacent first-order peaks in $|S({\bf k})|$, as illustrated in Fig.~\ref{fig:3}(b),
and for the $A_a = 0$ triangular lattice, $\theta_l \approx 60^\circ$.
The triangular lattice is slightly distorted by our perfectly square simulation box, which gives us the minor deviation from
the ideal value
$\theta_l = 60^\circ$.
In Fig.~\ref{fig:3}(c,d),
at $A_a = 0.039$
the vortices form a rhombic
lattice with $\theta_l \approx 76^{\circ}$.
At $A_a = 0.099$ in Fig.~\ref{fig:3}(e,f),
a square lattice
appears with $\theta_l = 90^{\circ}$.
In Fig.~\ref{fig:4} we plot the lattice angle $\theta_l$
versus $A_a$ for samples with $A_v = 1.0$, 1.5, 2.0, and $2.5$.
In all cases,
at low $A_a \lesssim 0.3$ the vortex ordering is triangular,
at intermediate $A_a$ a rhombic lattice structure appears,
and at large $A_a \gtrsim 0.5$ a square
VL emerges.
In Fig.~\ref{fig:5} we plot a structural phase diagram indicating where the triangular, rhombic, and square
VLs appear as a function of $A_a$ versus $A_v$.
The evolution of the VL symmetry is nearly independent of the value of $A_v$,
showing that the triangular-to-square transition is a robust feature of our MD simulations.

\begin{figure}
\includegraphics[width=3.5in]{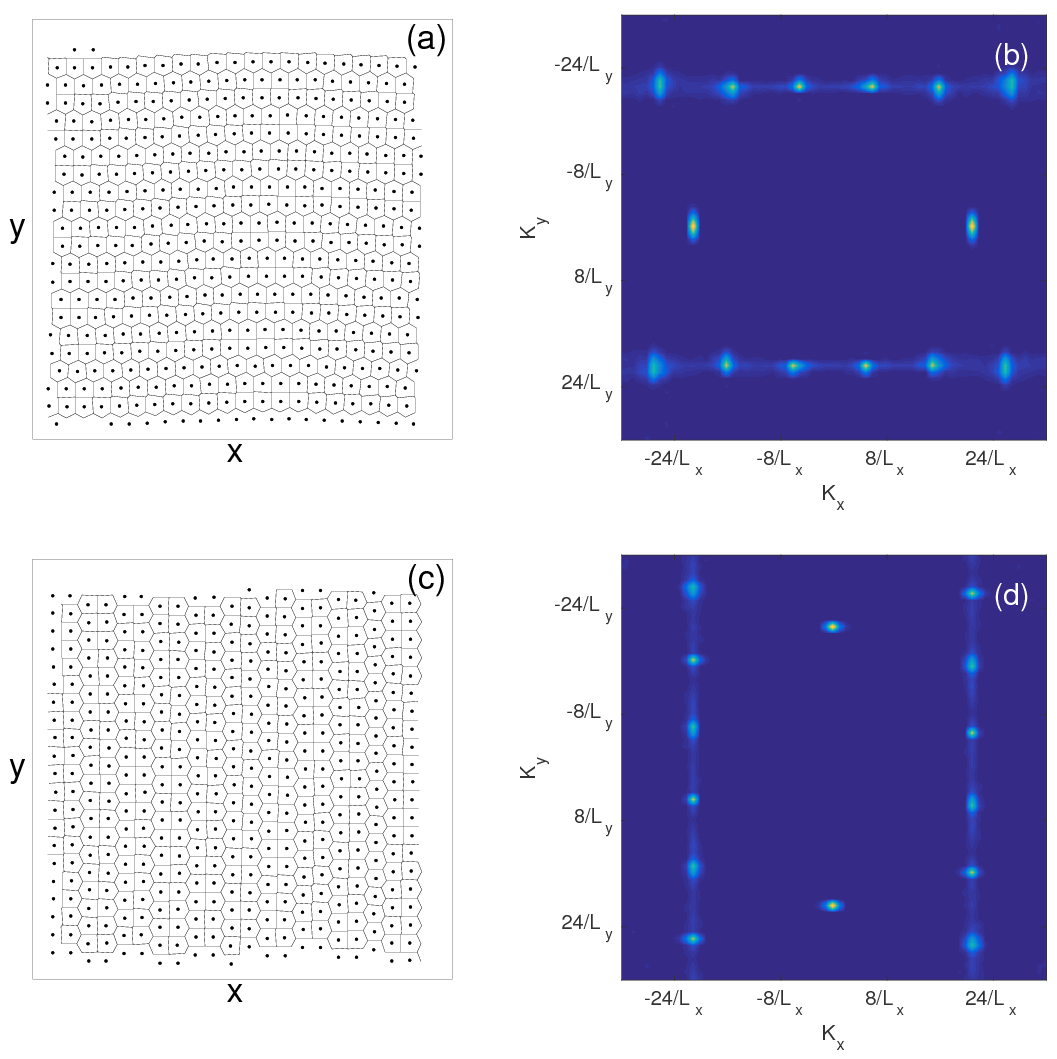}
\caption{Multi-$q$ Voronoi constructions for the real space vortex positions
  (a,c)
  and the corresponding structure factor $|S({\bf k})|$ (b,d)
in samples with
  $A_a = 0.04$ at
  $A_v = 2.0$ (a,b) and
  $A_v = 2.8$ (c,d).
  The Voronoi polygons indicate that there is a combination of square and triangular ordering in the form of an Archimedean tiling,
  producing  multiple peaks in $|S({\bf k})|$.
}
\label{fig:6}
\end{figure}

For some combinations of $A_a$ and $A_v$
what we term a multi-$q$ state appears
in which the vortices exhibit simultaneous square and triangular ordering.
Figure~\ref{fig:6}
shows both
the real space Voronoi polygons and the corresponding $|S({\bf k})|$
for multi-$q$ states in samples with $A_a = 0.04$.
For both $A_v=2.0$ in Fig.~\ref{fig:6}(a,b) and $A_v=2.8$ in Fig.~\ref{fig:6}(c,d) we find the same
combination of square and triangular ordering in the Voronoi polygons,
while $|S({\bf k})|$ exhibits multiple sets of peaks.  As illustrated
in Fig.~\ref{fig:6}, 
the multi-$q$ ordering can be oriented along either the $x$ or the $y$ direction.
The multi-$q$ VL structure closely resembles an
Archimedean tiling in which space is filled with a combination of square and triangular tiles.\cite{41}
Archimedean ordering of this type has also
been observed for colloidal assemblies driven over quasiperiodic substrates,
where it arises due to the competition between the ordering imposed by the substrate and the triangular ordering that minimizes the colloid-colloid interaction energy.\cite{42,43} 
In our system the competition responsible for producing this structure is between the square and triangular orderings favored by the anisotropy.    
We note that the multi-$q$ state only appears occasionally
in regions of $A_a$ and $A_v$ that are dominated by the rhombic state,
suggesting that it could be metastable.
Additionally, the 
two different orientations of the multi-$q$
state that we find indicate that in diffraction experiments
on macroscopic samples, domains of different orientations
will most likely coexist.
This would smear out the peaks in $|S({\bf k})|$, making it difficult to deconvolute the signal from an individual domain orientation.
As a result, local
probe imaging techniques
may be the best method for observing multi-$q$ states.

\begin{figure}
\includegraphics[width=3.5in]{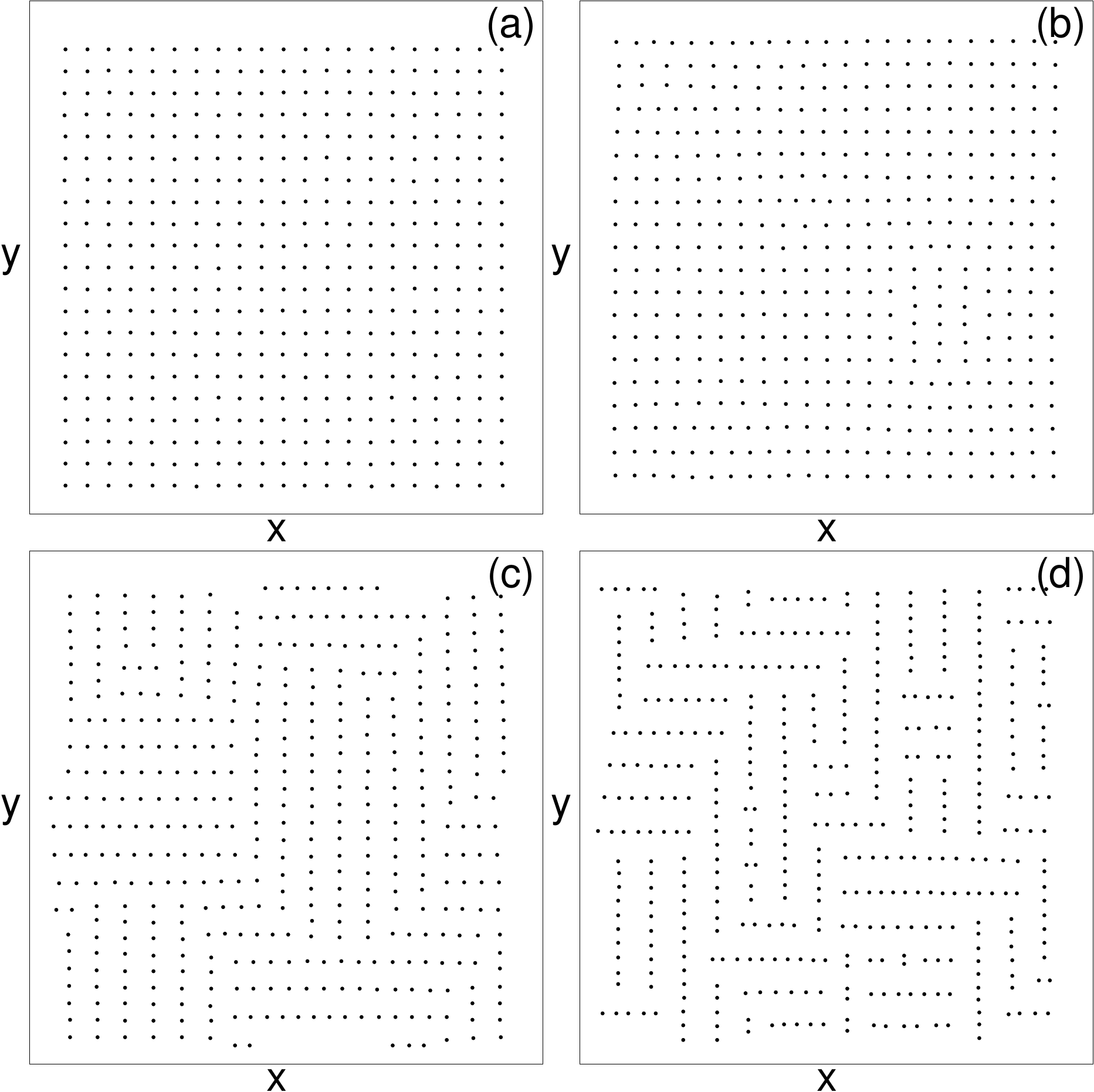}
\caption{Real space vortex configurations showing the emergence
  of chain states in systems with $A_v=2.0$.
    At $A_a = 0.1$ (a) the VL 
  is square.
  For $A_a = 0.6$ (b) the square lattice develops some local distortions and dislocations.
  At $A_a = 1.3$ (c) chain state domains appear,
  and these become more pronounced for $A_a = 3.0$ (d).
}
\label{fig:7}
\end{figure}

For anisotropies $A_a$ larger than those discussed above, we find
that the square lattice gradually transforms into domains of vortex chains.
This process is illustrated in Fig.~\ref{fig:7} for systems
with $A_v = 2.0$ where $A_a$ is increased from $A_a=0.1$ to $A_a=3.0$.
Although such levels of anisotropy may seem unphysically large,
there have been several  observations of vortex chain states,
including domains of chains in different borocarbides and in Sr$_2$RuO$_4$
at low fields.\cite{44,45,46}
While these observations are typically attributed to an attractive interaction
between vortices at intermediate range,
it may still be possible
to model these chain states by introducing strongly
anisotropic vortex-vortex interactions.

\section{Summary}
We
have introduced a model for vortices with anisotropic pairwise interactions,
  focusing on the case of four-fold asymmetry.
Using MD simulations we show that this model captures a transition from a triangular lattice at low anisotropy to a square VL
at high anisotropy,
with an intermediate rhombic phase.
We also find that in some cases a multi-$q$ state
with Archimedean ordering appears in which the vortices have both square and triangular local ordering.
For the highest anisotropy values, domains of vortex chains form.
Our model could be applied to study the dynamics near the VL
transitions,
where nonequilibrium phenomena can arise.\cite{47,48}
Additionally,
it can be generalized for higher order anisotropy in order to capture other types of symmetry and reorientation transitions in VLs.
\cite{49,50,51,52,53}.  
It is also possible to use the 
model with different isotropic pairwise interactions to investigate hexagonal to square transitions in other particle-based systems,
such as skyrmions or colloids with anisotropic interactions.     

\acknowledgments
We are grateful to D. Green, M. Lamichhane, X. Ma, D. McDermott and K. Newman for assistance and discussions.
This research was supported in part by the Notre Dame Center for Research Computing.
M.R.E. was supported by the U.S. Department of Energy, Office of Basic Energy Sciences, under Award No. DE-SC0005051.
This work was carried out under the auspices of the NNSA of the U.S. DoE at LANL under Contract No. DE-AC52-06NA25396.

\end{document}